\documentclass[sigconf]{acmart}

\usepackage{amsmath}
\usepackage{amsfonts}
\usepackage{algorithm}
\usepackage{algorithmic}
\usepackage{booktabs}
\usepackage{multirow}
\usepackage{lscape}
\usepackage{graphicx}
\usepackage{geometry}

\AtBeginDocument{%
  }

\copyrightyear{2024}
\acmYear{2024}
\setcopyright{acmlicensed}\acmConference[ICAIF '24]{5th ACM International
Conference on AI in Finance}{November 14--17, 2024}{Brooklyn, NY, USA}
\acmBooktitle{5th ACM International Conference on AI in Finance (ICAIF '24),
November 14--17, 2024, Brooklyn, NY, USA}
\acmDOI{10.1145/3677052.3698656}
\acmISBN{979-8-4007-1081-0/24/11}

\begin{document}

\title{Numin: Weighted-Majority Ensembles for Intraday Trading}


\author{Aniruddha Mukherjee}
\email{2205533@kiit.ac.in}
\affiliation{%
  \institution{Kalinga Institute of Industrial Technology}
  \city{Bhubaneswar}
  \country{India}
}

\author{Rekha Singhal}
\email{rekha.singhal@tcs.com}
\affiliation{%
  \institution{TCS Research}
  \city{New York City}
  \country{USA}
}

\author{Gautam Shroff}
\email{gautam.shroff@iiitd.ac.in}
\affiliation{%
  \institution{IIIT, Delhi}
  \city{New Delhi}
  \country{India}
}




\begin{abstract}
{We consider the application of machine learning
models for short-term intra-day trading in equities. We envisage
a scenario wherein
machine learning models 
are submitted by independent data scientists 
to predict discretised ten-candle returns every five minutes,
in response to five-minute candlestick data provided to them in near 
real-time. An ensemble model combines these multiple models 
via a weighted-majority algorithm. The weights of each
model are dynamically updated based on the performance
of each model, and can also be used to reward model owners.
Each model's performance is evaluated according to two different 
metrics over a recent time window:
In addition
to accuracy, we also consider a `utility' metric that is 
a proxy for a model's potential profitability
under a particular trading strategy.
We present experimental results on real intra-day data that show that
our weighted-majority ensemble techniques show improved 
accuracy as well as utility over any of the individual models,
especially using the utility metric to dynamically re-weight
models over shorter time-windows.
}
\end{abstract}

\ccsdesc[500]{Computing methodologies~Ensemble methods}
\ccsdesc[300]{Applied computing~Forecasting}
\ccsdesc[100]{Applied computing~Economics}
\ccsdesc[300]{Mathematics of computing~Time series analysis}
\ccsdesc[300]{Information systems~Data stream mining}

\keywords{Stock Prediction, Intra-day trading, Dynamic model weighting, Weighted-majority algorithm, Ensemble models}


\maketitle

\section{Introduction}

In recent years, the integration of advanced machine learning techniques with financial trading has garnered significant interest within both academia and industry. The advent of competitions such as Numerai \cite{numerai} has highlighted the potential for collaborative and competitive forecasting to drive innovation in financial markets. Numerai, a hedge fund, harnesses the power of a global community of data scientists to predict 20-day stock returns, rewarding top-performing models with cryptocurrency. Inspired by this model, our research explores the application of similar ensemble-learning strategies to intra-day trading in the Indian equities markets, towards predicting 10-candle returns using data sampled every five minutes. 

This paper introduces Numin, a framework designed to aggregate predictions from multiple models to optimize trading decisions in near real-time. Our approach leverages a variant of the (Dynamic) Weighted Majority Algorithm \cite{kotler} to dynamically assign weights to each model based on their historical performance, ensuring that the most reliable predictions are given the highest consideration. The core objective is to determine the most effective method for weighting and aggregating predictions from a diverse set of models to achieve superior trading outcomes.

We employ eight distinct predictive models, each utilizing unique strategies to forecast the 10-candle return of selected stocks. To dynamically adjust the weights of these models, we utilize historical accuracy as well as a \textit{utility} metric \cite{nabar2023conservative}. Each metric is calculated over a predefined window, giving more weight to models that have demonstrated consistent performance. The utility metric is a proxy for the profitability of a predictions, 
under the assumption that long/short positions are taken only when models predict extreme values, and are exited exactly after 10 candles.

Our experiments explore various weighting mechanisms, including dynamically updated weights based on different recent performance windows and exponentially moving averages, to suggest an optimal aggregation strategy. By continuously updating these weights, our system adapts to changing market conditions, aiming to rapidly identify the most accurate predictive models, with the goal of making trades based on weighted-majority predictions that are better than using any individual model.
Further, weights of each model can be used to dynamically reward model builders, in a manner similar to that used by Numerai \cite{numerai}.

Our experiments with real-life (5 minute candle) streaming data suggest that using \textbf{utility}, rather than accuracy as the scoring mechanism
with which to dynamically weight models, over a \textbf{short} (25 minute) window, may be optimal as well as useful (i.e., resulting in positive average utility), beating the `average' performance of individual models \footnote{Note that the average of any metric across all models is unachievable in practice.}.
Importantly, the performance of each model is estimated in a manner simulating actual trading under realistic assumptions.

\section{Background}

\subsection{Ensemble Learning}

Ensemble learning combines predictions from multiple models. Boosting, bagging, and adaptive mixture of experts are some common ensemble-learning
techniques. The weighted majority algorithm~\cite{LITTLESTONE1994212wma} and its variants~\cite{kotler} and~\cite{liu2022adaptation} dynamically weight different
models based on their recent performance. A weighted-majority (WMA) prediction is made using weighted majority voting, i.e., each target class gets
a score computed as the sum of the weights of all models in the ensemble that predict that class. The class getting the most score is chosen as
the prediction for the WMA ensemble. Different weighted majority algorithms ensembles differ on how the weights are dynamically adjusted, e.g.
the original WMA~\cite{LITTLESTONE1994212wma} algorithm adjusted the weights multiplicatively with each prediction as soon as its true label was revealed,
and had provable mistake bounds.
Other variants use model predictions across a window, and/or alternative metrics and updation rules. 
We use both accuracy and a trading specific metric called \textit{utility} over various window sizes in our experiments.

\subsection{Numerai}

Numerai \cite{numerai} is a pioneering hedge fund that leverages collective intelligence of a global community of data scientists to make stock market predictions. Participants submit their models' predictions for a 20-day return on a set of anonymised stocks. The Numerai meta-model aggregates these predictions along with dynamically adjusted model weights to rank the collection of stocks, using which the hedge fund takes long-short market-neutral positions.
Participants whose models consistently perform well are rewarded in Numerai's proprietary cryptocurrency, which can be converted into real money. 
As compared to Numerai, we collect real-time predictions (for five-minute candles) in an intra-day setting, and assume direct positions (rather
than an arbitrage-based long-short strategy).

\section{Methods: Algorithms \& Models}

In this section, we describe our variant of the Weighted Majority Algorithm (WMA) employed to aggregate predictions of models, as well as the 8 models used themselves.


\subsection{Weighted Majority Algorithm (WMA)}
\label{sec3:wma_methods}
The Weighted Majority Algorithm (WMA) is utilized to aggregate predictions from $n$ experts, each providing forecasts for the 10-candle return of stocks. In a high-frequency trading scenario where data is received at 5-minute intervals, determining the optimal model in real-time poses a significant challenge. The efficacy of each model can only be evaluated in hindsight which is why making a choice for the ``best expert'' of the $n$ in real time challenging. The WMA addresses this issue by weighing each expert's predictions on the basis of their performance on a metric $\phi$ over a historical window of recent data. The ``final prediction'' is calculated by a weighed majority vote from all experts based on their corresponding weights. This is done ensure that the ensemble prediction is the best possible one given the information available at that time regarding each model's performance in the near past.

As mentioned, the algorithm functions by assigning weights to each expert based on their historical performance over a recent time window. As new data continuously arrives, these weights are dynamically adjusted, enabling the algorithm to prioritize the models that have demonstrated superior performance in the recent past. This adaptive weighting voting mechanism allows the system to leverage the varying strengths of each expert as market conditions evolve, thus enhancing the overall predictive \textcolor{black}{prowess}.

By incorporating WMA, the system effectively manages the ensemble of expert predictions, thereby optimizing the decision-making process in a continuously changing market environment. The algorithm’s ability to update weights in real-time ensures that the aggregated prediction remains robust and reflective of the most current market trends.

\begin{algorithm}
\caption{\textcolor{black}{WMA} Weight Initialization Using Historical Candle Data with EMA (``\textcolor{black}{training mode}'')}
\begin{algorithmic}[1]
\REQUIRE $n$ models, $m$ tickers (i.e., stock symbols), evaluation metric $\phi$, min window size $\mu$, max window size $\lambda$, train rounds $r_{{\text{start}}_\text{train}}$ to $r_{\text{end}} = 75$
\STATE Initialize weights \textcolor{black}{for $n$ models:} $w_j = \frac{1}{n}$ for $j = 1, 2, \ldots, n$
\FOR{round $r = r_{{\text{start}}_\text{train}}$ to $r_{\text{end}}$}
    \FOR{ticker $t = 1$ to $m$}
        \STATE Append $(r, t)$ data to window $\text{win}$
        \IF{$|\text{win}| > \lambda$}
            \STATE Remove the oldest round from $\text{win}$
        \ENDIF

        \STATE $p_{j,t} = M_j.\text{predict}(r, t)$

        \IF{$|\text{win}| \geq \mu$}
             \STATE $s_{j,t}^{(r)} = \phi(\{p_{j,t}^{(r')}\}_{r' \in \text{win}}, \{T_t^{(r')}\}_{r' \in \text{win}})$ 
             
            \COMMENT{Scored using $\phi$ comparing $p_{j,t}$ with ground-truth $T_t$ over rounds $r$ in win}
        \ENDIF
    \ENDFOR

    \IF{$|\text{win}| \geq \mu$}
        \STATE Compute average and normalized scores for each model $M_j$ \textcolor{black}{across tickers}:
        \[
        s_j^{(r)} = \frac{1}{m} \sum_{t=1}^m s_{j,t}, \quad \tilde{s}_j^{(r)} = \frac{s_j^{(r)}}{\sum_{k=1}^n s_k^{(r)}}
        \]
        \STATE Calculate the EMA weight update with $\alpha = \frac{2}{|\text{win}| + 1}$:
        \STATE $w_j^{(r)} = \alpha \cdot \tilde{s}_j^{(r)} + (1 - \alpha) \cdot w_j^{(r-1)}$ \COMMENT{EMA weight update}
    \ELSE
        \STATE $w_j^{(r)} = \frac{1}{n}$
    \ENDIF
\ENDFOR
\STATE Finalize the weights after the last train round:
\[
w_j = w_j^{(r_{\text{end}})} \quad \text{for} \quad j = 1, 2, \ldots, n
\]
\end{algorithmic}
\end{algorithm}


We sample equity candles at 5-minute intervals from near the start of the financial day nearing closing, resulting in 75 `rounds' of data for which each
expert model needs to make predictions that are combined using WMA to produce a final recommendation for each round as it arrives. For a live trading day, we also concatenate the data of the previous financial day: Thus rounds 0 to 74 are data sampled at 5-min intervals during the immediately preceding financial day; the current day's data starts from round 75 and continues until round 149. Note that data values are all normalized using the closing
price of the first candle of the day, i.e., round 75; as a result, data values for all stocks (`tickers') fall within the same range independent of
actual prices; further, the ten-candle return is also binned using a separate dataset into five ordinal classes (this normalisation procedure is detailed Section 4\textcolor{black}{)}.

Our system has two `modes': in `training' mode, using the data available from the previous day, we ``warm-up'' the weights for each model. We define a window of variable length, i.e., $\mu$ rounds must exist historically to constitute a usable window. Until then we weight each model equally at $w_j = \frac{1}{n}$ for $j = 1, 2, \ldots n$ and predict using standard majority voting. Once there are more than $\mu$ rounds, we use a metric $\phi$ to determine how well each model performed historically in these rounds, using which the models are assigned weights based on their performance\footnote{Note that all ground truth labels for the previous day,
i.e., ten-candle returns are available from the start of the current day, while during the current day these labels arrive with a delay of 10 steps.}.
After each round model weights are updated using an exponential moving average to retain a portion of the historical trend and enable more stable weight updates by smoothing out the fluctuations. This initialization of weight and it's update mechanism in training mode is described in Algorithm 1.

The second mode is the ``inference'' mode or the live, online, continuous mode for the current day.
During inference mode, each model makes a predictions for current round which are combined using weighted majority as explained earlier, using current model weights
\footnote{Note that at the start of the current day, i.e., round 75, weights post training mode are available.}; Each model has a weight. Now, each model makes a prediction to predict one of 5 classes. Each class is initially given a weight of 0. When $\text{model}_{i}$ predicts class $j$, the weight assigned to class $j$ is incremented by the weight of $\text{model}_{i}$ calculated in the previous round. The class with the maximum weight at the end is selected to be the ``WMA'' prediction.

Next weight updates are computed based on a performance metric $\phi$ (e.g., accuracy or utility, defined below), which necessitates a ground truth value, which can only be obtained 10 rounds prior to the current round. Therefore $\phi$ is computed for each model using a window of length of $\mu$ ending 10 rounds prior to the current round (note that this window may spill over into the previous day for the initial rounds 75-84 of the current day). Weights are 
normalized, so that they sum to one, and then updated using an exponential moving average to dampen oscillations. 
The inference mode of the model is detailed in Algorithm 2.

\begin{algorithm}
\caption{Inference using WMA and EMA (Test Phase)}
\begin{algorithmic}[1]
\REQUIRE $n$ models, $m$ tickers, evaluation metric $\phi$, max window size $\lambda$, test rounds $r_{\text{start}} = 75$ to $r_{\text{end}} = 149$, train start round $r_{{\text{start}}_\text{train}}$
\STATE Initialize weights $w_j = \frac{1}{n}$ for $j = 1, 2, \ldots, n$
\STATE Initialize data structures: $\mathcal{P} \gets \emptyset$
\STATE Initialize window $\text{win}$ with available data from rounds $r_{{\text{start}}_\text{train}}$ to 65 
\FOR{round $r = r_{\text{start}}$ to $r_{\text{end}}$}
    \IF{$r > 75$}
        \STATE Append data of round $r-10$ to window $\text{win}$ \COMMENT{Ground truth for $r-10$ becomes available}
        \IF{$|\text{win}| > \lambda$}
            \STATE Remove the oldest round from $\text{win}$
        \ENDIF
    \ENDIF
    \FOR{ticker $t = 1$ to $m$}

        \STATE $p_{j,t}^{(r)} = M_j.\text{predict}(r, t)$

        \STATE $s_{j,t}^{(r)} = \phi(\{p_{j,t}^{(r')}\}_{r' \in \text{win}}, \{T_t^{(r')}\}_{r' \in \text{win}})$ \COMMENT{Score using $\phi$ comparing predictions with ground-truths over rounds in win}
    \ENDFOR
    \STATE Compute average and normalized scores for each model $M_j$:
    \[
    s_j^{(r)} = \frac{1}{m} \sum_{t=1}^m s_{j,t}^{(r)}, \quad \tilde{s}_j^{(r)} = \frac{s_j^{(r)}}{\sum_{k=1}^n s_k^{(r)}}
    \]
    \STATE Calculate the EMA weight update with $\alpha = \frac{2}{|\text{win}| + 1}$:
    \STATE $w_j^{(r)} = \alpha \cdot \tilde{s}_j^{(r)} + (1 - \alpha) \cdot w_j^{(r-1)}$ \COMMENT{EMA weight update}
    \STATE Initialize class voting weights $\mathcal{V} \gets \{c_k: 0 \mid k \in [0, 4]\}$
    \FOR{each model $M_j$}
        \STATE $\mathcal{V}[c_{p_{j,t}^{(r)}}] \gets \mathcal{V}[c_{p_{j,t}^{(r)}}] + w_j^{(r-1)}$
    \ENDFOR
    \STATE $\mathcal{P}[r][t] \gets \arg\max_k \mathcal{V}[c_k]$
\ENDFOR
\STATE Finalize the weights and output final predictions: $\mathcal{P}$
\end{algorithmic}
\end{algorithm}

\subsection{Predictive Models}

We utilize eight models in our experiments, each leveraging different machine learning techniques to forecast the 10-candle return value. These models are described below:

\subsubsection{Categorical Convolutional Neural Network (CNN)}
A window size of 40 was selected. Using a sliding window of 40 rounds, OHLC candlestick plots were created. The $y_{max}$ and $y_{min}$ of the plots were $+2\sigma_{High_n}$ and $-2\sigma_{Low_n}$, $\sigma$ being the standard deviation of the data used to pre-train the model. The CNN architecture consists of two convolutional layers with 16 and 32 filters (3x3), each followed by a max-pooling layer (2x2). The output is flattened and passed through three dense layers with 128, 64, and 16 units (ReLU activation). The final dense layer has 5 units, representing the class probabilities. A categorical cross-entropy loss function is utilized along with the Adam optimizer with a learning rate of $0.01$. The model was pre-trained with a large dataset of historical data. At test time, the model creates an image, and using it's pre-training knowledge only, it makes a prediction. It is worth mentioning that the y-axis scale remains the same throughout.

\subsubsection{Convolutional Neural Network (CNN) Regressor}
This model is very similar one above, with the images being fed in the same and the architecture also being exactly identical with the exception of the final layer which is a Dense layer with 1 unit corresponding to a single continuous value representing 10-candle return. Mean squared error is used as the loss function along with the Adam optimizer with a learning rate of $0.01$. This continuous predicted value is binned into one of 5 categories using the same bin-edges to convert the continuous 10-candle-return into one of 5 classes.

\subsubsection{Long Short-Term Memory (LSTM) Categorical}
The LSTM model processes sequences of 40 data points to capture temporal dependencies in stock price movements. The architecture consists of an input layer, two LSTM layers with 50 units each, and a dense output layer with 5 units representing the class probabilities. The first LSTM layer returns sequences, while the second returns only the last output. The model is trained using negative log-likelihood loss on a large dataset of historical stock price data. The final prediction is the class label of the 10-candle return, categorized into 5 classes.

\subsubsection{Long Short-Term Memory (LSTM) Regression}
The LSTM model operates on sequences of 40 data points, capturing temporal dependencies and patterns in the stock price movements. The model shares the same architecture with the previously described model, with the exception of the output layer. In this LSTM model, the dense output layer has 1 unit corresponding to a single continuous value representing the final prediction which is a single value for the normalized 10-candle return, which is binned into one of 5 categories using the same bin edges to convert the continuous return into discrete classes. The model is trained using negative log-likelihood loss on a large dataset of historical stock price data.

\subsubsection{Continual Multi-Layer Perceptron (MLP)}
This model employs an online learning approach, continuously updating its knowledge based on recent data points. The architecture consists of an MLP with customizable hidden layers (default is one layer with 100 neurons) and uses a sliding window technique to manage input data.
The model maintains a memory of recent data points, with a configurable window size (default 15). For each prediction, it is fine-tuned on the most recent window of data, allowing it to adapt to changing market conditions. The MLP uses backpropagation with a specified learning rate (default 0.01) to update its weights.
The model predicts one of 5 classes for the 10-candle return, where each candle represents a 5-minute interval. This effectively forecasts the market direction 50 minutes into the future.
During inference, the model processes the input features through its trained neural network to produce a prediction. Continual learning using fine-tuning allows the model to evolve its understanding of market patterns over time, potentially capturing short-term trends and anomalies. Pre-trained weights for the MLP, obtained by training on the same large historical dataset as for previous models, are loaded at the start of each day.

\subsubsection{Mixture of Experts (MoE)}
This model employs a Mixture of Experts (MoE) architecture for financial market prediction. It consists of 10 expert networks, each a MLP with four fully connected layers, and a gating network that dynamically weights the experts' outputs. The model uses ReLU activations and dropout for regularization. During training, it leverages transfer learning by initializing with pre-trained weights and then fine-tunes these on recent data, excluding the 10 most recent samples since labels are not available for these. This approach allows the model to balance historical knowledge with current market dynamics. For inference, the gating network assigns weights to each expert's prediction based on input features, using which the 5-class categorical prediction of the ten-candle return
is output. This MoE is also pre-trained on the same historical dataset as for the earlier models.

\subsubsection{AutoEncoder Test-Time Adapter}
This model employs a Test-Time Adaptation approach combined with an Autoencoder and Classifier architecture for financial market prediction. The core components include an autoencoder for feature extraction and a separate classifier for prediction. The model uses a unique training strategy where it continuously adapts during inference. The autoencoder, with a symmetrical structure (35 → 64 → 32 → 16 → 32 → 64 → 35 neurons), learns to reconstruct input features, while the classifier (16 → 16 → 5 neurons) makes predictions based on the encoded representations. During inference, the model first updates its autoencoder weights using the current input data, then uses the updated encoder to generate features for classification. This adaptive approach allows the model to continually refine its understanding of market conditions. The output is a 5-class probability distribution, likely corresponding to different market movement scenarios. The strength of this architecture lies in its ability to dynamically adjust to changing market patterns in real-time, potentially capturing short-term market dynamics more effectively than static models.

\subsubsection{K-Means Clustering}
This model implements a Mixture of Experts (MoE) approach combined with K-Means clustering for financial market prediction. The architecture consists of multiple expert models (MLP classifiers) and a K-Means clustering algorithm for expert selection. The model uses pre-trained experts trained on different clusters obtained
using K-Means clustering on historical data (the same dataset as used for earlier models). During inference, the system dynamically selects an expert based on the input data's proximity to cluster centroids and the experts' historical performance. The selection process uses a weighted probability scheme that considers both the distance to cluster centroids and the experts' recent prediction accuracy. The model adapts its expert selection weights based on recent prediction outcomes, reducing the influence of poorly performing experts. This adaptive mechanism allows the model to adjust to changing market conditions by favoring more successful experts over time. The output is a 5-class prediction of the 10-candle return.

\section{Materials, Data, and Features}
Data was sampled from the Indian equities market at 5-minute intervals starting from the 20th of September 2023 until the 13th of June 2024. We used equity price data captured as five-minute candles in the standard open, high, low, close, and volume (OHLCV) form. Data for each ticker (stock) is normalized for each day by the close price of the first candle, resulting in each day starting with a normalized close price of one, with remaining values throughout the day recorded in multiples of this value.

To enhance the predictive capabilities of our models, we engineered a diverse set of features from the raw stock data. These features include normalized opening, highest, lowest, closing prices, and volume of stocks within each round (\textit{$\text{Open}_n$, $\text{High}_n$, $\text{Low}_n$, $\text{Close}_n$, $\text{Volume}_n$}), which
are the normalised values using the close of the first candle of the current day for OHLC and the historical average volume for each ticker to compute $\text{Volume}_n$. Simple Moving Averages (computed using $\text{Close}_n$, i.e., normalised values) over 10 and 20 rounds (\textit{\textit{$\text{SMA}_{10}$} and \textit{$\text{SMA}_{20}$}}) help smooth out price data and highlight trends over different time periods. The Relative Strength Index calculated over 14 rounds (\textit{$\text{RSI}_{14}$}) indicates overbought or oversold conditions. Bollinger Bands (\textit{$\text{BBL}_{\text{5, 2.0}}$, $\text{BBM}_{\text{5, 2.0}}$, $\text{BBU}_{\text{5, 2.0}}$}) capture price range and volatility, with Bollinger Bandwidth and Percent B values (\textit{$\text{BBB}_{\text{5, 2.0}}$ and $\text{BBP}_{\text{5, 2.0}}$}) providing additional insights. The Moving Average Convergence Divergence (\textit{$\text{MACD}_{\text{12, 26, 9}}$}), its Histogram (\textit{$\text{MACDh}_{\text{12, 26, 9}}$}), and Signal line (\textit{$\text{MACDs}_{\text{12, 26, 9}}$}) help identify trend direction, strength, and potential reversals. The Volume Weighted Average Price (\textit{$\text{VWAP}_{\text{D}}$}) represents the average price a stock has traded at throughout the day, weighted by volume. Momentum indicators such as \textit{$\text{MOM}_{\text{30}}$} show the rate of change in closing prices, while the Chande Momentum Oscillator (\textit{$\text{CMO}_{\text{14}}$}) measures price momentum and identifies overbought or oversold conditions. Differences between the highest and lowest normalized prices (\textit{$\text{High}_{\text{n}}$-$\text{Low}_{\text{n}}$}) and between normalized opening and closing prices (\textit{$\text{Open}_{\text{n}}$-$\text{Close}_{\text{n}}$}) capture volatility and intraday price changes, respectively. The difference between the 20-round and 10-round Simple Moving Averages (\textit{$\text{SMA}_{\text{20}}$-$\text{SMA}_{\text{10}}$}) provides insights into momentum and potential reversal points. Slopes of the normalized closing prices over three, five, and ten rounds (\textit{${\text{Close}_{\text{n}}}_{\text{slope\_3}}$, ${\text{Close}_{\text{n}}}_{\text{slope\_5}}$, ${\text{Close}_{\text{n}}}_{\text{slope\_10}}$}) capture trend direction and strength over different periods. 

Finally, a number of `changelen' parameters are computed: each captures the maximum number of of recent rounds for which a value has been increasing or decreasing. So a changelen value of 3 means the value has been increasing for the 3 past rounds; similarly, a -3 would mean the value was decreasing for the past 3 rounds. Changelen values help identify persistent trends or reversals, and are computed for normalized opening, highest, lowest, and closing prices, as well as the differences between highest-lowest, opening-closing prices, and $\text{SMA}_{\text{20}}$-$\text{SMA}_{\text{10}}$.

The sampled data was split into three parts. Part one contained data from 14 days ($\text{21}^{\text{st}}$ September 2023 to $\text{12}^{\text{th}}$ October 2023) concatenated with a large volume of data sampled from 2022. This data was used to calculate the mean and standard deviation of the normalized data (as each day is normalized by dividing all the OHLC values by the close price at the start of the new day at round 75) for each feature. The second part contained data from the $\text{13}^{\text{th}}$ of October 2023 until the $\text{22}^{\text{nd}}$ of March 2024 (88 days). The features were z-normalized using the means ($\mu$) and standard deviations ($\sigma$) calculated from the prior historical data of part one, yielding a standard normal distribution for each feature with $\mu$ and $\sigma$ hovering around 0 and 1 respectively. The third part contains data from the $\text{2}^{\text{nd}}$ of April 2024 to the $\text{13}^{\text{th}}$ of June 2024 (29 days), which was also z-normalized in a similar manner as the second part, yielding ($\mu$, $\sigma$) $\sim$ (0,1). Normalization helps in handling the scale differences among features, ensuring that each feature contributes equally to the model's predictions. We use a subset of the data from the 2nd of May 2024 till the 14th of May 2024 (covering a 10-day span) to perform our experiments.

\subsection{Structure of Data}

As mentioned in Section~\ref{sec3:wma_methods}, the data is structured into rounds, with each round containing the new data that was sampled in the 5 minutes that passed by since the last round. Rounds 0 to 74 represent the previous trading day (this is immediately available and obviously does not need to be sampled), rounds 75 onwards till round 149 correspond to the current trading day. This division facilitates a clear distinction between training and testing phases, allowing our models to warm up on historical data as described in Algorithm 1 before making predictions in the real-time scenario. The training phase utilizes data from rounds prior to the current trading day (i.e. rounds 0 to 74), enabling the models to learn from historical patterns. This setup ensures that the WMA system has enough historical data to initialize the weights of the models. The testing phase described in Algorithm 2 involves making predictions based on the latest incoming data, with continuous updates and adjustments to model weights based on performance metrics over a span of time (or a window).

We computed the target values as the difference between the closing price 50 minutes into the future (round $r+10$) and the current closing price, normalized by the close price when the markets open for trading. These returns were then discretized into five categories: 0, 0.25, 0.5, 0.75, and 1. The bins were chosen in a manner that ensured a roughly equal distribution among the 5 classes. For evaluation, we measured accuracy and utility. Utility (defined below) assess the model's profitability. For instance, if the actual target is 1, indicating a price rise, but the model predicts 0, suggesting a price fall, this misprediction results in a loss when we exit the long/short position taken 50 minutes prior. This framework allows us to determine if the model correctly anticipates profitable positions.

\subsection{Simulating Trading for Test-time Metrics}
\label{real-sim-trade}
To get the final ``accuracy'' / ``utility'' we simulate a  real-time trading strategy: If the model being evaluated predicts an extreme value, e.g., a 4 or a 0, indicating extreme positive or negative 10-candle return, a long or short position is appropriately taken (else no position is taken); this position exited only after 10-candles.
Thus, model predictions in the interim, while computed for the purposes of WMA re-weighting, are not used in practice and should ideally be ignored for computing test-time metrics. Thus, we mask test-data for each model, including the WMA ensembles, to simulate trading as follows:
When a 0 or 4 is predicted we mask the next 10 rounds for a ticker $t$ when calculating the accuracy or utility. Note that this leads to
varying support for each model as each model's predictions are different and so are the masks used for it. Total utility, which is not dimensionless, is thus normalised by the test support for each model post masking, resulting in comparable values, i.e., representing the `average utility per trade recommended'\footnote{Note that we have ignored transaction costs in this simulated metric; however, these can easily be incorporated as well}.


\section{Experiments}
This section describes the experiments conducted to evaluate the performance of various models and weight update strategies using the Weighted Majority Algorithm (WMA). Each experiment aims to determine the effectiveness of using different metrics (accuracy and utility) with which models are compared for updating weights,
over different window lengths. Note that weight update take place using an exponential moving average (EMA) with smoothing factor \(\alpha = \) \textcolor{black}{$\frac{2}{len(\text{win}) + 1}$}\footnote{\textcolor{black}{Note the defintion https://www.investopedia.com/terms/e/ema.asp}}
\[
\text{EMA}_t = \alpha \cdot \text{Value}_t + (1 - \alpha) \cdot \text{EMA}_{t-1}
\]
A higher \(\alpha\) favors recent values, while a lower \(\alpha\) remembers more of the past.

\begin{figure*}
    \centering
    \includegraphics[width=\linewidth]{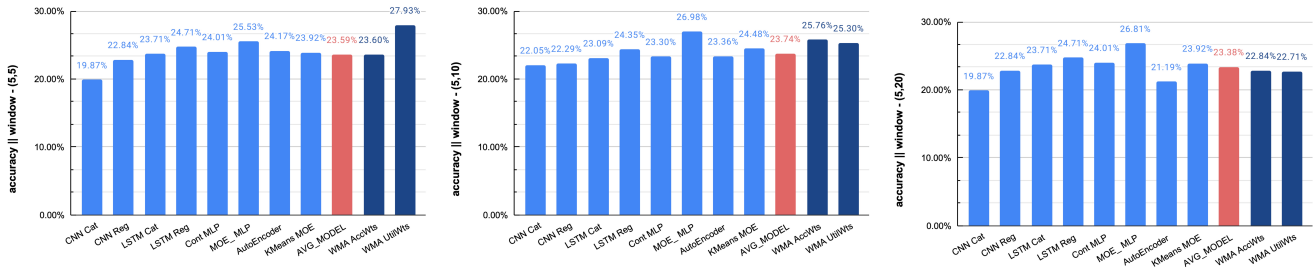}
    \caption{\textcolor{black}{Accuracies of the 8 models along with an ``AVG\_MODEL'' and the two WMA methods over 3 different window sizes.}}
    \Description{A plot showing accuracies of eight models, along with AVG\_MODEL and two WMA methods, compared over three different window sizes.}
    \label{fig:enter-label}
\end{figure*}

\subsubsection{Accuracy Metric Updating Weights}
Accuracy measures the proportion of correct predictions made by each model over a specified window:
\[
\text{Accuracy} = \frac{\text{Number of Correct Predictions}}{\text{Total Number of Predictions}}
\]
Using accuracy as the $\phi$ metric for weighting the models means that models that have made more correct predictions in the recent window will have higher weights in the next window.

\subsubsection{Utility as a Metric for Weighting the Models}

Utility is a proxy for the potential profitability of a model's predictions. It is calculated based assuming a trading strategy where a long/short position 
is taken in response to extreme predictions (e.g. 4 or \textcolor{black}{0}) and held over a 50-minute (10 rounds) period. In this context, the utility is defined as follows, where the rows represent the actual outcomes and the columns represent the predicted outcomes. The element $U_{ij}$ represents the utility awarded
to a trading strategy that predicted \( j \) when the actual outcome turned out to be \( i \). 

\begin{figure*}[h]
    \centering
    \includegraphics[width=\linewidth]{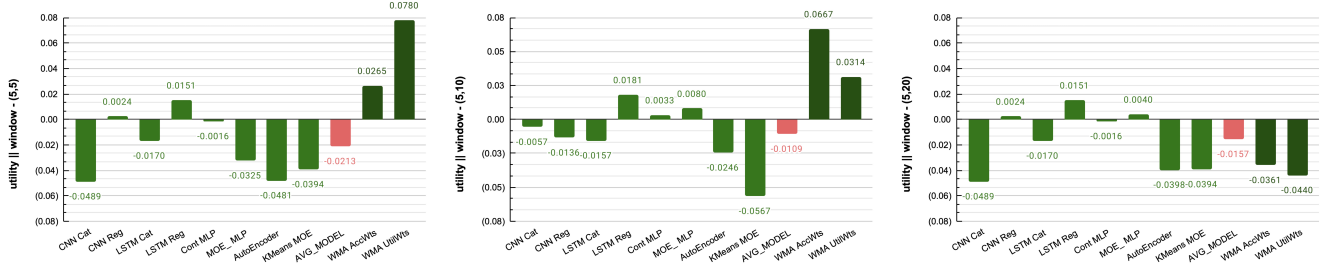}
    \caption{\textcolor{black}{Utility of the 8 models along with an ``AVG\_MODEL'' and the two WMA methods over 3 different window sizes.}}
    \Description{Utility of 8 models, including AVG_MODEL and WMA methods, over different window sizes.}
    \label{fig:util-results}
\end{figure*}

\[
U = \begin{bmatrix}
2 & 0 & 0 & 0 & -2 \\
1 & 0 & 0 & 0 & -1 \\
0 & 0 & 0 & 0 & 0 \\
-1 & 0 & 0 & 0 & 1 \\
-2 & 0 & 0 & 0 & 2
\end{bmatrix}
\]

The utility matrix $U$ is structured such that positive values represent beneficial predictions, while negative values indicate detrimental predictions. For instance, predicting the actual outcome correctly (diagonal elements) yields a utility of 2 for extreme predictions (4 or 0), but 0 for other predictions 
in response to which no position is taken.

\begin{figure}[h]
    \centering
    \includegraphics[width=\linewidth]{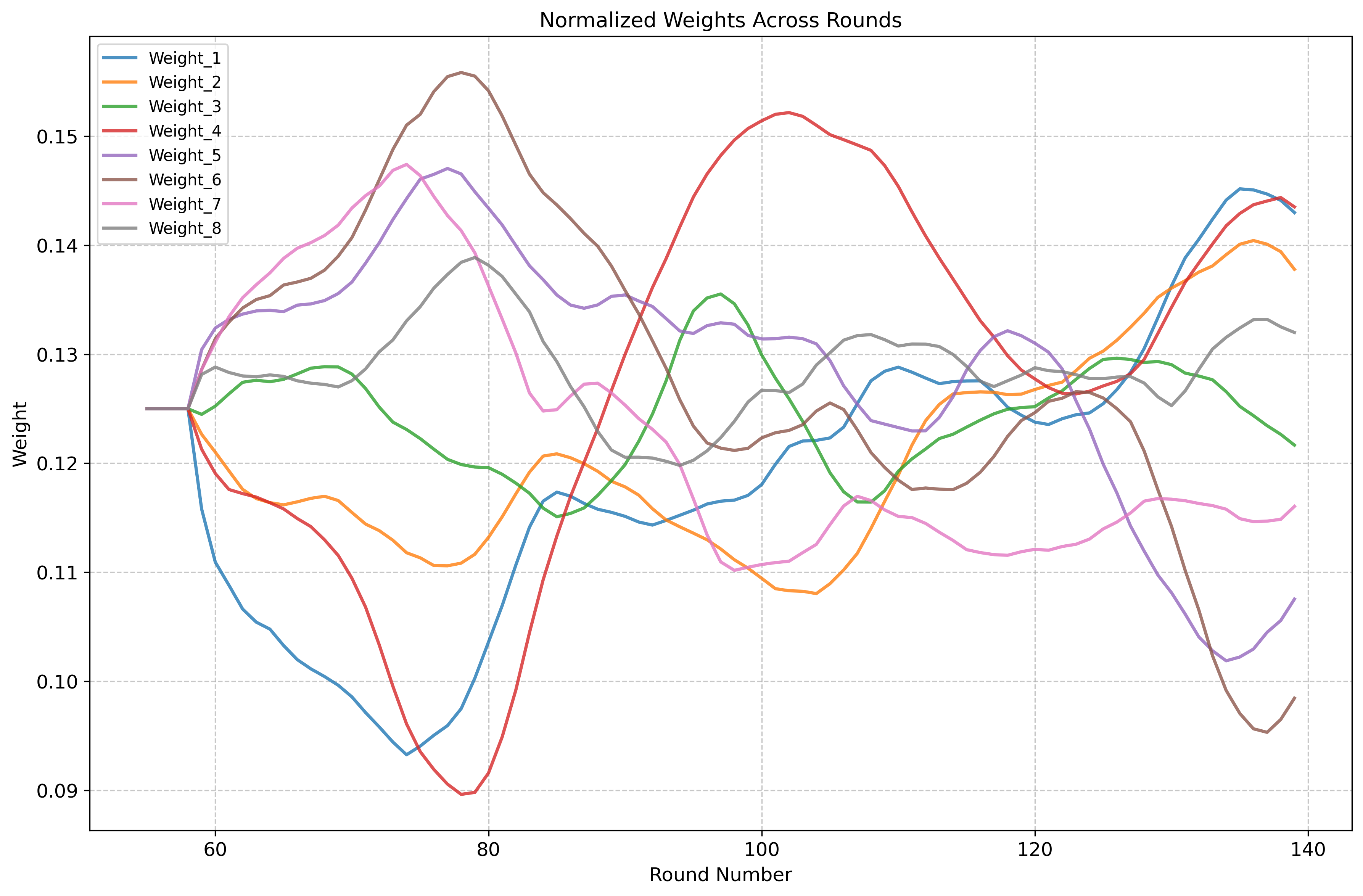}
    \Description{Temporal evolution of normalized weights for eight models using EMA and model accuracy as $\phi$ on May 8th.}
    \caption{Temporal evolution of normalized weights for eight models using Exponential Moving Average (EMA), a (5,10) window, and $\phi$ set to model accuracy, observed on May 8th yielding a 0.18 Utility.}
    \label{fig:weight-updates}
\end{figure}

On the other hand incorrect predictions (off-diagonal elements), result in negative utility
(but \textit{only} for extreme predictions) if the actual outcome
is directionally opposite to that predicted, e.g. an outcome of 1 or 0 when a 4 is predicted, or an outcome of 3 or 4 when a 0 is predicted. However, if
the outcome is directional similar still differs from the extreme value predicted, e.g. outcome of 4 when predicting a 4, we still award a (smaller) positive utility to this prediction since some profit albeit lower may result nevertheless. (Note that transaction costs are not incorporated in this definition, but can easily be introduced by adjusting the entries; the only conceptual difference that this would lead to is that an outcome of 2 when predicting 0 or 4 would still result in a small negative utility instead of zero).

Note that when utility is used as a metric for weighting, the weights are adjusted based on the potential profits derived from the models' predictions. Models that contribute more to the overall utility will be given higher weights, ensuring that the ensemble benefits from the most practically useful models. 

Of course we report utility of test predictions even when accuracies are used to weight models, as well as vice versa.

\subsubsection{Results}

Table~\ref{tab:acc} summarizes the accuracies of each model, accuracies of WMA using accuracy and utility as $\phi$ and a hypothetical ``AVG MODEL'' obtained by averaging the accuracies of the 8 standalone models. 

Table~\ref{tab:util-tab} presents the average utility per round (prediction) for each of the 8 models averaged over the 5 days (2, 3, 8, 10, 14th of May) used to conduct our experiments. We also report the utility when WMA is employed with both accuracy and utility as $\phi$. 

Note that as explained earlier in \textcolor{black}{Section \ref{real-sim-trade}}, test results are computed simulating real trading by masking rows based on the assumed trading
strategy where no trades are made for a ticker when holding an open position in it.

Performance metrics are computed for differently sized windows of recent data. Each reported metric is obtained by averaging across all days in the test set.

\begin{table}[h]
\centering
\caption{Accuracy reported for each method averaged over the 5 days over which experiments were performed}
\label{tab:acc}
\resizebox{\columnwidth}{!}{%
\begin{tabular}{@{}l|ll|ll|ll@{}}
\toprule
\multicolumn{1}{r}{Window Sizes} & \multicolumn{2}{c}{(5,5)} & \multicolumn{2}{c}{(5,10)} & \multicolumn{2}{c}{(5,20)} \\ \midrule
Methods     & accuracy & std  & accuracy & std  & accuracy & std  \\ \midrule
CNN Cat     & 19.87\%  & 0.06 & 22.05\%  & 0.04 & 19.87\%  & 0.06 \\
CNN Reg     & 22.84\%  & 0.05 & 22.29\%  & 0.05 & 22.84\%  & 0.05 \\
LSTM Cat    & 23.71\%  & 0.03 & 23.09\%  & 0.04 & 23.71\%  & 0.03 \\
LSTM Reg    & 24.71\%  & 0.06 & 24.35\%  & 0.06 & 24.71\%  & 0.06 \\
Cont MLP    & 24.01\%  & 0.03 & 23.30\%  & 0.04 & 24.01\%  & 0.03 \\
MOE\_ MLP   & 25.53\%  & 0.03 & \textbf{26.98}\%  & 0.03 & \textbf{26.81}\%  & 0.02 \\
AutoEncoder & 24.17\%  & 0.07 & 23.36\%  & 0.02 & 21.19\%  & 0.01 \\
KMeans MOE  & 23.92\%  & 0.05 & 24.48\%  & 0.04 & 23.92\%  & 0.05 \\ \midrule
AVG\_MODEL  & 23.59\%  & 0.01 & 23.74\%  & 0.02 & 23.38\%  & 0.02 \\
WMA AccWts  & 23.60\%  & 0.05 & 25.76\%  & 0.03 & 22.84\%  & 0.03 \\
WMA UtilWts & \textbf{27.93}\%  & 0.07 & 25.30\%  & 0.03 & 22.71\%  & 0.05 \\ \bottomrule
\end{tabular}%
}
\end{table}

\begin{table}[h]
\centering
\caption{Utility reported for each method averaged over the 5 days over which experiments were performed}
\label{tab:util-tab}
\resizebox{\columnwidth}{!}{%
\begin{tabular}{@{}lrlrlrl@{}}
\toprule
\multicolumn{1}{r}{Window Sizes} & \multicolumn{2}{c}{(5,5)}                      & \multicolumn{2}{c}{(5,10)}                  & \multicolumn{2}{c}{(5,20)} \\ \midrule
\multicolumn{1}{l|}{Methods} &
  \multicolumn{1}{c}{utility} &
  \multicolumn{1}{c|}{std} &
  \multicolumn{1}{c}{utility} &
  \multicolumn{1}{c|}{std} &
  \multicolumn{1}{c}{utility} &
  \multicolumn{1}{c}{std} \\ \midrule
\multicolumn{1}{l|}{CNN Cat}     & -0.0489         & \multicolumn{1}{l|}{-0.0057} & -0.0057         & \multicolumn{1}{l|}{0.16} & -0.0489           & 0.16   \\
\multicolumn{1}{l|}{CNN Reg}     & 0.0024          & \multicolumn{1}{l|}{-0.0136} & -0.0136         & \multicolumn{1}{l|}{0.02} & 0.0024            & 0.02   \\
\multicolumn{1}{l|}{LSTM Cat}    & -0.0170         & \multicolumn{1}{l|}{-0.0157} & -0.0157         & \multicolumn{1}{l|}{0.01} & -0.0170           & 0.01   \\
\multicolumn{1}{l|}{LSTM Reg}    & 0.0151          & \multicolumn{1}{l|}{0.0181}  & 0.0181          & \multicolumn{1}{l|}{0.03} & \textbf{0.0151}   & 0.04   \\
\multicolumn{1}{l|}{Cont MLP}    & -0.0016         & \multicolumn{1}{l|}{0.0033}  & 0.0033          & \multicolumn{1}{l|}{0.04} & -0.0016           & 0.04   \\
\multicolumn{1}{l|}{MOE\_ MLP}   & -0.0325         & \multicolumn{1}{l|}{0.0080}  & 0.0080          & \multicolumn{1}{l|}{0.06} & 0.0040            & 0.06   \\
\multicolumn{1}{l|}{AutoEncoder} & -0.0481         & \multicolumn{1}{l|}{-0.0246} & -0.0246         & \multicolumn{1}{l|}{0.09} & -0.0398           & 0.06   \\
\multicolumn{1}{l|}{KMeans MOE}  & -0.0394         & \multicolumn{1}{l|}{-0.0567} & -0.0567         & \multicolumn{1}{l|}{0.06} & -0.0394           & 0.07   \\ \midrule
\multicolumn{1}{l|}{AVG\_MODEL}  & -0.0213         & \multicolumn{1}{l|}{-0.0109} & -0.0109         & \multicolumn{1}{l|}{0.03} & -0.0157           & 0.03   \\
\multicolumn{1}{l|}{WMA AccWts}  & 0.0265          & \multicolumn{1}{l|}{0.0667}  & \textbf{0.0667} & \multicolumn{1}{l|}{0.08} & -0.0361           & 0.07   \\
\multicolumn{1}{l|}{WMA UtilWts} & \textbf{0.0780} & \multicolumn{1}{l|}{0.0314}  & 0.0314          & \multicolumn{1}{l|}{0.08} & -0.0440           & 0.10   \\ \bottomrule
\end{tabular}%
}
\end{table}

\section{Discussion}
The Weighted Majority Algorithm (WMA) is utilized to aggregate predictions from $n$ experts, each providing forecasts for the 10-candle return of stocks. This approach aims to identify the best-performing expert in real-time. The results indicate that WMA generally outperforms the ``AVG\_MODEL'', a hypothetical model with accuracy or utility equivalent to the average of the accuracies or utilities of the ensembled models. Notably, aside from a few exceptions, both the accuracy-weighted and utility-weighted versions of WMA consistently demonstrate superior utility compared to individual models.

The impact of window size on model performance was also examined. It was observed that smaller window sizes perform better. A window size of ($\mu$ = 5, $\lambda$ = 5) using utility as the weighting mechanism performs well. However, extending the window to $\lambda$ = 20 rounds (approximately two hours into the past) appears to enhance WMA's accuracy. In contrast, for utility, a window size of $\lambda$ = 10 rounds (50 minutes into the past) produces optimal results, with utility diminishing as the window size increases to 20 rounds. 

Figure \ref{fig:weight-updates} illustrates how model weights change over the course of a trading day when using a small window size: During the warming-up period, for the first $w$ rounds of the previous day
weights are equal, after which the change fairly rapidly as the ensemble adjusts to favor those models that are doing better as measured
by the chosen metric (utility here) over the given window (5,10).

Note that the average utility per round as reported for smaller window sizes, i.e., 5,5 and 5,10, is positive for the WMA methods, 
indicating potentially profitable trades on the average, even though most of the individual models show negative average utility, 
indicating the WMA is judiciously favoring the right model at the right time. 

These findings suggest that WMA is an effective method for aggregating expert predictions, consistently outperforming the average model
and resulting in potential profitability even when each individual model may not be profitable on the average. Moreover, the choice of window size and weighting mechanism can further refine the performance, with shorter windows favoring utility and longer windows favoring accuracy. This nuanced understanding of WMA's performance characteristics can inform the development of more effective forecasting models in financial markets.

\begin{table*}[]
\centering
\caption{Complete experimental results for all 5 days using all 11 approaches}
\label{tab:exp_res}
\resizebox{\textwidth}{!}{%
\begin{tabular}{|cc|ccccccc||ccccccc||ccccccc|}
\hline
\multicolumn{2}{|c|}{Window Size} & \multicolumn{7}{c|}{(5,5)} & \multicolumn{7}{c|}{(5,10)} & \multicolumn{7}{c|}{(5,20)} \\ \hline
\multicolumn{2}{|c|}{Date} & \multicolumn{1}{c|}{02-May} & \multicolumn{1}{c|}{03-May} & \multicolumn{1}{c|}{08-May} & \multicolumn{1}{c|}{10-May} & \multicolumn{1}{c|}{14-May} & \multicolumn{1}{c|}{\multirow{3}{*}{\begin{tabular}[c]{@{}c@{}}STDDEV\\ across\\ days\end{tabular}}} & \multirow{3}{*}{\begin{tabular}[c]{@{}c@{}}AVG\_ACC\\ across\\ days\end{tabular}} & \multicolumn{1}{c|}{02-May} & \multicolumn{1}{c|}{03-May} & \multicolumn{1}{c|}{08-May} & \multicolumn{1}{c|}{10-May} & \multicolumn{1}{c|}{14-May} & \multicolumn{1}{c|}{\multirow{3}{*}{\begin{tabular}[c]{@{}c@{}}STDDEV\\ across\\ days\end{tabular}}} & \multirow{3}{*}{\begin{tabular}[c]{@{}c@{}}AVG\_ACC\\ across\\ days\end{tabular}} & \multicolumn{1}{c|}{02-May} & \multicolumn{1}{c|}{03-May} & \multicolumn{1}{c|}{08-May} & \multicolumn{1}{c|}{10-May} & \multicolumn{1}{c|}{14-May} & \multicolumn{1}{c|}{\multirow{3}{*}{\begin{tabular}[c]{@{}c@{}}STDDEV\\ across\\ days\end{tabular}}} & \multirow{3}{*}{\begin{tabular}[c]{@{}c@{}}AVG\_ACC\\ across\\ days\end{tabular}} \\ \cline{1-7} \cline{10-14} \cline{17-21}
\multicolumn{2}{|c|}{Num of Rounds} & \multicolumn{1}{c|}{149} & \multicolumn{1}{c|}{149} & \multicolumn{1}{c|}{149} & \multicolumn{1}{c|}{149} & \multicolumn{1}{c|}{149} & \multicolumn{1}{c|}{} &  & \multicolumn{1}{c|}{149} & \multicolumn{1}{c|}{149} & \multicolumn{1}{c|}{149} & \multicolumn{1}{c|}{149} & \multicolumn{1}{c|}{149} & \multicolumn{1}{c|}{} &  & \multicolumn{1}{c|}{149} & \multicolumn{1}{c|}{149} & \multicolumn{1}{c|}{149} & \multicolumn{1}{c|}{149} & \multicolumn{1}{c|}{149} & \multicolumn{1}{c|}{} &  \\ \cline{1-7} \cline{10-14} \cline{17-21}
\multicolumn{2}{|c|}{Num of Tickers} & \multicolumn{1}{c|}{5} & \multicolumn{1}{c|}{6} & \multicolumn{1}{c|}{8} & \multicolumn{1}{c|}{8} & \multicolumn{1}{c|}{7} & \multicolumn{1}{c|}{} &  & \multicolumn{1}{c|}{5} & \multicolumn{1}{c|}{6} & \multicolumn{1}{c|}{8} & \multicolumn{1}{c|}{8} & \multicolumn{1}{c|}{7} & \multicolumn{1}{c|}{} &  & \multicolumn{1}{c|}{5} & \multicolumn{1}{c|}{6} & \multicolumn{1}{c|}{8} & \multicolumn{1}{c|}{8} & \multicolumn{1}{c|}{7} & \multicolumn{1}{c|}{} &  \\ \hline
\multicolumn{1}{|c|}{\multirow{2}{*}{CNN Cat}} & Accuracy & \multicolumn{1}{c|}{10.77\%} & \multicolumn{1}{c|}{26.56\%} & \multicolumn{1}{c|}{19.70\%} & \multicolumn{1}{c|}{22.86\%} & \multicolumn{1}{c|}{19.44\%} & \multicolumn{1}{c|}{\textbf{0.06}} & \textbf{19.87\%} & \multicolumn{1}{c|}{18.82\%} & \multicolumn{1}{c|}{29.41\%} & \multicolumn{1}{c|}{19.70\%} & \multicolumn{1}{c|}{22.86\%} & \multicolumn{1}{c|}{19.44\%} & \multicolumn{1}{c|}{\textbf{0.04}} & \textbf{22.05\%} & \multicolumn{1}{c|}{10.77\%} & \multicolumn{1}{c|}{26.56\%} & \multicolumn{1}{c|}{19.70\%} & \multicolumn{1}{c|}{22.86\%} & \multicolumn{1}{c|}{19.44\%} & \multicolumn{1}{c|}{\textbf{0.06}} & \textbf{19.87\%} \\ \cline{2-23} 
\multicolumn{1}{|c|}{} & Utility & \multicolumn{1}{c|}{-0.2154} & \multicolumn{1}{c|}{0.1406} & \multicolumn{1}{c|}{0.0303} & \multicolumn{1}{c|}{-0.2071} & \multicolumn{1}{c|}{0.0069} & \multicolumn{1}{c|}{\textbf{0.16}} & \textbf{-0.0489} & \multicolumn{1}{c|}{-0.0941} & \multicolumn{1}{c|}{0.2353} & \multicolumn{1}{c|}{0.0303} & \multicolumn{1}{c|}{-0.2071} & \multicolumn{1}{c|}{0.0069} & \multicolumn{1}{c|}{\textbf{0.16}} & \textbf{-0.0057} & \multicolumn{1}{c|}{-0.2154} & \multicolumn{1}{c|}{0.1406} & \multicolumn{1}{c|}{0.0303} & \multicolumn{1}{c|}{-0.2071} & \multicolumn{1}{c|}{0.0069} & \multicolumn{1}{c|}{\textbf{0.16}} & \textbf{-0.0489} \\ \hline
\multicolumn{1}{|c|}{\multirow{2}{*}{CNN Reg}} & Accuracy & \multicolumn{1}{c|}{29.59\%} & \multicolumn{1}{c|}{27.27\%} & \multicolumn{1}{c|}{16.83\%} & \multicolumn{1}{c|}{18.66\%} & \multicolumn{1}{c|}{21.86\%} & \multicolumn{1}{c|}{\textbf{0.05}} & \textbf{22.84\%} & \multicolumn{1}{c|}{26.98\%} & \multicolumn{1}{c|}{27.13\%} & \multicolumn{1}{c|}{16.83\%} & \multicolumn{1}{c|}{18.66\%} & \multicolumn{1}{c|}{21.86\%} & \multicolumn{1}{c|}{\textbf{0.05}} & \textbf{22.29\%} & \multicolumn{1}{c|}{29.59\%} & \multicolumn{1}{c|}{27.27\%} & \multicolumn{1}{c|}{16.83\%} & \multicolumn{1}{c|}{18.66\%} & \multicolumn{1}{c|}{21.86\%} & \multicolumn{1}{c|}{\textbf{0.05}} & \textbf{22.84\%} \\ \cline{2-23} 
\multicolumn{1}{|c|}{} & Utility & \multicolumn{1}{c|}{0.0000} & \multicolumn{1}{c|}{0.0379} & \multicolumn{1}{c|}{0.0096} & \multicolumn{1}{c|}{-0.0075} & \multicolumn{1}{c|}{-0.0279} & \multicolumn{1}{c|}{\textbf{0.02}} & \textbf{0.0024} & \multicolumn{1}{c|}{-0.0476} & \multicolumn{1}{c|}{0.0053} & \multicolumn{1}{c|}{0.0096} & \multicolumn{1}{c|}{-0.0075} & \multicolumn{1}{c|}{-0.0279} & \multicolumn{1}{c|}{\textbf{0.02}} & \textbf{-0.0136} & \multicolumn{1}{c|}{0.0000} & \multicolumn{1}{c|}{0.0379} & \multicolumn{1}{c|}{0.0096} & \multicolumn{1}{c|}{-0.0075} & \multicolumn{1}{c|}{-0.0279} & \multicolumn{1}{c|}{\textbf{0.02}} & \textbf{0.0024} \\ \hline
\multicolumn{1}{|c|}{\multirow{2}{*}{LSTM Cat}} & Accuracy & \multicolumn{1}{c|}{22.67\%} & \multicolumn{1}{c|}{20.83\%} & \multicolumn{1}{c|}{21.60\%} & \multicolumn{1}{c|}{25.68\%} & \multicolumn{1}{c|}{27.78\%} & \multicolumn{1}{c|}{\textbf{0.03}} & \textbf{23.71\%} & \multicolumn{1}{c|}{17.78\%} & \multicolumn{1}{c|}{22.63\%} & \multicolumn{1}{c|}{21.60\%} & \multicolumn{1}{c|}{25.68\%} & \multicolumn{1}{c|}{27.78\%} & \multicolumn{1}{c|}{\textbf{0.04}} & \textbf{23.09\%} & \multicolumn{1}{c|}{22.67\%} & \multicolumn{1}{c|}{20.83\%} & \multicolumn{1}{c|}{21.60\%} & \multicolumn{1}{c|}{25.68\%} & \multicolumn{1}{c|}{27.78\%} & \multicolumn{1}{c|}{\textbf{0.03}} & \textbf{23.71\%} \\ \cline{2-23} 
\multicolumn{1}{|c|}{} & Utility & \multicolumn{1}{c|}{1.9933} & \multicolumn{1}{c|}{1.9792} & \multicolumn{1}{c|}{2.0000} & \multicolumn{1}{c|}{1.9797} & \multicolumn{1}{c|}{1.9630} & \multicolumn{1}{c|}{\textbf{0.01}} & \textbf{1.9830} & \multicolumn{1}{c|}{-0.0056} & \multicolumn{1}{c|}{-0.0158} & \multicolumn{1}{c|}{0.0000} & \multicolumn{1}{c|}{-0.0203} & \multicolumn{1}{c|}{-0.0370} & \multicolumn{1}{c|}{\textbf{0.01}} & \textbf{-0.0157} & \multicolumn{1}{c|}{-0.0067} & \multicolumn{1}{c|}{-0.0208} & \multicolumn{1}{c|}{0.0000} & \multicolumn{1}{c|}{-0.0203} & \multicolumn{1}{c|}{-0.0370} & \multicolumn{1}{c|}{\textbf{0.01}} & \textbf{-0.0170} \\ \hline
\multicolumn{1}{|c|}{\multirow{2}{*}{LSTM Reg}} & Accuracy & \multicolumn{1}{c|}{25.00\%} & \multicolumn{1}{c|}{32.39\%} & \multicolumn{1}{c|}{15.61\%} & \multicolumn{1}{c|}{25.08\%} & \multicolumn{1}{c|}{25.47\%} & \multicolumn{1}{c|}{\textbf{0.06}} & \textbf{24.71\%} & \multicolumn{1}{c|}{22.63\%} & \multicolumn{1}{c|}{32.97\%} & \multicolumn{1}{c|}{15.61\%} & \multicolumn{1}{c|}{25.08\%} & \multicolumn{1}{c|}{25.47\%} & \multicolumn{1}{c|}{\textbf{0.06}} & \textbf{24.35\%} & \multicolumn{1}{c|}{25.00\%} & \multicolumn{1}{c|}{32.39\%} & \multicolumn{1}{c|}{15.61\%} & \multicolumn{1}{c|}{25.08\%} & \multicolumn{1}{c|}{25.47\%} & \multicolumn{1}{c|}{\textbf{0.06}} & \textbf{24.71\%} \\ \cline{2-23} 
\multicolumn{1}{|c|}{} & Utility & \multicolumn{1}{c|}{-0.0067} & \multicolumn{1}{c|}{-0.0208} & \multicolumn{1}{c|}{0.0000} & \multicolumn{1}{c|}{-0.0203} & \multicolumn{1}{c|}{-0.0370} & \multicolumn{1}{c|}{\textbf{0.01}} & \textbf{-0.0170} & \multicolumn{1}{c|}{-0.0158} & \multicolumn{1}{c|}{0.0769} & \multicolumn{1}{c|}{0.0074} & \multicolumn{1}{c|}{0.0127} & \multicolumn{1}{c|}{0.0094} & \multicolumn{1}{c|}{\textbf{0.03}} & \textbf{0.0181} & \multicolumn{1}{c|}{-0.0278} & \multicolumn{1}{c|}{0.0739} & \multicolumn{1}{c|}{0.0074} & \multicolumn{1}{c|}{0.0127} & \multicolumn{1}{c|}{0.0094} & \multicolumn{1}{c|}{\textbf{0.04}} & \textbf{0.0151} \\ \hline
\multicolumn{1}{|c|}{\multirow{2}{*}{Cont MLP}} & Accuracy & \multicolumn{1}{c|}{21.33\%} & \multicolumn{1}{c|}{29.52\%} & \multicolumn{1}{c|}{22.91\%} & \multicolumn{1}{c|}{24.43\%} & \multicolumn{1}{c|}{21.88\%} & \multicolumn{1}{c|}{\textbf{0.03}} & \textbf{24.01\%} & \multicolumn{1}{c|}{18.82\%} & \multicolumn{1}{c|}{28.45\%} & \multicolumn{1}{c|}{22.91\%} & \multicolumn{1}{c|}{24.43\%} & \multicolumn{1}{c|}{21.88\%} & \multicolumn{1}{c|}{\textbf{0.04}} & \textbf{23.30\%} & \multicolumn{1}{c|}{21.33\%} & \multicolumn{1}{c|}{29.52\%} & \multicolumn{1}{c|}{22.91\%} & \multicolumn{1}{c|}{24.43\%} & \multicolumn{1}{c|}{21.88\%} & \multicolumn{1}{c|}{\textbf{0.03}} & \textbf{24.01\%} \\ \cline{2-23} 
\multicolumn{1}{|c|}{} & Utility & \multicolumn{1}{c|}{1.9800} & \multicolumn{1}{c|}{1.9277} & \multicolumn{1}{c|}{2.0223} & \multicolumn{1}{c|}{2.0305} & \multicolumn{1}{c|}{2.0313} & \multicolumn{1}{c|}{\textbf{0.04}} & \textbf{1.9984} & \multicolumn{1}{c|}{-0.0118} & \multicolumn{1}{c|}{-0.0560} & \multicolumn{1}{c|}{0.0223} & \multicolumn{1}{c|}{0.0305} & \multicolumn{1}{c|}{0.0313} & \multicolumn{1}{c|}{\textbf{0.04}} & \textbf{0.0033} & \multicolumn{1}{c|}{-0.0200} & \multicolumn{1}{c|}{-0.0723} & \multicolumn{1}{c|}{0.0223} & \multicolumn{1}{c|}{0.0305} & \multicolumn{1}{c|}{0.0313} & \multicolumn{1}{c|}{\textbf{0.04}} & \textbf{-0.0016} \\ \hline
\multicolumn{1}{|c|}{\multirow{2}{*}{MOE\_ MLP}} & Accuracy & \multicolumn{1}{c|}{23.85\%} & \multicolumn{1}{c|}{21.32\%} & \multicolumn{1}{c|}{28.41\%} & \multicolumn{1}{c|}{29.36\%} & \multicolumn{1}{c|}{24.69\%} & \multicolumn{1}{c|}{\textbf{0.03}} & \textbf{25.53\%} & \multicolumn{1}{c|}{26.67\%} & \multicolumn{1}{c|}{22.84\%} & \multicolumn{1}{c|}{28.41\%} & \multicolumn{1}{c|}{29.36\%} & \multicolumn{1}{c|}{27.61\%} & \multicolumn{1}{c|}{\textbf{0.03}} & \textbf{26.98\%} & \multicolumn{1}{c|}{24.17\%} & \multicolumn{1}{c|}{24.26\%} & \multicolumn{1}{c|}{28.65\%} & \multicolumn{1}{c|}{29.36\%} & \multicolumn{1}{c|}{27.61\%} & \multicolumn{1}{c|}{\textbf{0.02}} & \textbf{26.81\%} \\ \cline{2-23} 
\multicolumn{1}{|c|}{} & Utility & \multicolumn{1}{c|}{-0.0278} & \multicolumn{1}{c|}{0.0739} & \multicolumn{1}{c|}{0.0074} & \multicolumn{1}{c|}{0.0127} & \multicolumn{1}{c|}{0.0094} & \multicolumn{1}{c|}{\textbf{0.04}} & \textbf{0.0151} & \multicolumn{1}{c|}{0.0667} & \multicolumn{1}{c|}{-0.0062} & \multicolumn{1}{c|}{-0.0909} & \multicolumn{1}{c|}{0.0092} & \multicolumn{1}{c|}{0.0613} & \multicolumn{1}{c|}{\textbf{0.06}} & \textbf{0.0080} & \multicolumn{1}{c|}{0.0667} & \multicolumn{1}{c|}{-0.0441} & \multicolumn{1}{c|}{-0.0730} & \multicolumn{1}{c|}{0.0092} & \multicolumn{1}{c|}{0.0613} & \multicolumn{1}{c|}{\textbf{0.06}} & \textbf{0.0040} \\ \hline
\multicolumn{1}{|c|}{\multirow{2}{*}{AE}} & Accuracy & \multicolumn{1}{c|}{36.25\%} & \multicolumn{1}{c|}{19.77\%} & \multicolumn{1}{c|}{22.22\%} & \multicolumn{1}{c|}{21.74\%} & \multicolumn{1}{c|}{20.90\%} & \multicolumn{1}{c|}{\textbf{0.07}} & \textbf{24.17\%} & \multicolumn{1}{c|}{25.45\%} & \multicolumn{1}{c|}{26.50\%} & \multicolumn{1}{c|}{22.22\%} & \multicolumn{1}{c|}{21.74\%} & \multicolumn{1}{c|}{20.90\%} & \multicolumn{1}{c|}{\textbf{0.02}} & \textbf{23.36\%} & \multicolumn{1}{c|}{21.31\%} & \multicolumn{1}{c|}{19.77\%} & \multicolumn{1}{c|}{22.22\%} & \multicolumn{1}{c|}{21.74\%} & \multicolumn{1}{c|}{20.90\%} & \multicolumn{1}{c|}{\textbf{0.01}} & \textbf{21.19\%} \\ \cline{2-23} 
\multicolumn{1}{|c|}{} & Utility & \multicolumn{1}{c|}{1.9750} & \multicolumn{1}{c|}{2.0000} & \multicolumn{1}{c|}{1.9596} & \multicolumn{1}{c|}{1.8696} & \multicolumn{1}{c|}{1.9552} & \multicolumn{1}{c|}{\textbf{0.05}} & \textbf{1.9519} & \multicolumn{1}{c|}{0.1182} & \multicolumn{1}{c|}{-0.0256} & \multicolumn{1}{c|}{-0.0404} & \multicolumn{1}{c|}{-0.1304} & \multicolumn{1}{c|}{-0.0448} & \multicolumn{1}{c|}{\textbf{0.09}} & \textbf{-0.0246} & \multicolumn{1}{c|}{0.0164} & \multicolumn{1}{c|}{0.0000} & \multicolumn{1}{c|}{-0.0404} & \multicolumn{1}{c|}{-0.1304} & \multicolumn{1}{c|}{-0.0448} & \multicolumn{1}{c|}{\textbf{0.06}} & \textbf{-0.0398} \\ \hline
\multicolumn{1}{|c|}{\multirow{2}{*}{KMeans MOE}} & Accuracy & \multicolumn{1}{c|}{15.94\%} & \multicolumn{1}{c|}{25.68\%} & \multicolumn{1}{c|}{24.30\%} & \multicolumn{1}{c|}{24.07\%} & \multicolumn{1}{c|}{29.59\%} & \multicolumn{1}{c|}{\textbf{0.05}} & \textbf{23.92\%} & \multicolumn{1}{c|}{18.56\%} & \multicolumn{1}{c|}{25.88\%} & \multicolumn{1}{c|}{24.30\%} & \multicolumn{1}{c|}{24.07\%} & \multicolumn{1}{c|}{29.59\%} & \multicolumn{1}{c|}{\textbf{0.04}} & \textbf{24.48\%} & \multicolumn{1}{c|}{15.94\%} & \multicolumn{1}{c|}{25.68\%} & \multicolumn{1}{c|}{24.30\%} & \multicolumn{1}{c|}{24.07\%} & \multicolumn{1}{c|}{29.59\%} & \multicolumn{1}{c|}{\textbf{0.05}} & \textbf{23.92\%} \\ \cline{2-23} 
\multicolumn{1}{|c|}{} & Utility & \multicolumn{1}{c|}{-0.0200} & \multicolumn{1}{c|}{-0.0723} & \multicolumn{1}{c|}{0.0223} & \multicolumn{1}{c|}{0.0305} & \multicolumn{1}{c|}{0.0313} & \multicolumn{1}{c|}{\textbf{0.04}} & \textbf{-0.0016} & \multicolumn{1}{c|}{-0.0515} & \multicolumn{1}{c|}{-0.0471} & \multicolumn{1}{c|}{-0.0467} & \multicolumn{1}{c|}{-0.1481} & \multicolumn{1}{c|}{0.0102} & \multicolumn{1}{c|}{\textbf{0.06}} & \textbf{-0.0567} & \multicolumn{1}{c|}{0.0145} & \multicolumn{1}{c|}{-0.0270} & \multicolumn{1}{c|}{-0.0467} & \multicolumn{1}{c|}{-0.1481} & \multicolumn{1}{c|}{0.0102} & \multicolumn{1}{c|}{\textbf{0.07}} & \textbf{-0.0394} \\ \hline
\multicolumn{1}{|c|}{\multirow{2}{*}{\textbf{\begin{tabular}[c]{@{}c@{}}AVERAGE \\ MODEL\end{tabular}}}} & \textbf{Accuracy} & \multicolumn{1}{c|}{\textbf{23.17\%}} & \multicolumn{1}{c|}{\textbf{25.42\%}} & \multicolumn{1}{c|}{\textbf{21.45\%}} & \multicolumn{1}{c|}{\textbf{23.98\%}} & \multicolumn{1}{c|}{\textbf{23.95\%}} & \multicolumn{1}{c|}{\textbf{0.01}} & \textbf{23.59\%} & \multicolumn{1}{c|}{\textbf{21.96\%}} & \multicolumn{1}{c|}{\textbf{26.98\%}} & \multicolumn{1}{c|}{\textbf{21.45\%}} & \multicolumn{1}{c|}{\textbf{23.98\%}} & \multicolumn{1}{c|}{\textbf{24.32\%}} & \multicolumn{1}{c|}{\textbf{0.02}} & \textbf{23.74\%} & \multicolumn{1}{c|}{\textbf{21.35\%}} & \multicolumn{1}{c|}{\textbf{25.79\%}} & \multicolumn{1}{c|}{\textbf{21.48\%}} & \multicolumn{1}{c|}{\textbf{23.98\%}} & \multicolumn{1}{c|}{\textbf{24.32\%}} & \multicolumn{1}{c|}{\textbf{0.02}} & \textbf{23.38\%} \\ \cline{2-23} 
\multicolumn{1}{|c|}{} & \textbf{Utility} & \multicolumn{1}{c|}{\textbf{0.7098}} & \multicolumn{1}{c|}{\textbf{0.7583}} & \multicolumn{1}{c|}{\textbf{0.7565}} & \multicolumn{1}{c|}{\textbf{0.7110}} & \multicolumn{1}{c|}{\textbf{0.7415}} & \multicolumn{1}{c|}{\textbf{0.02}} & \textbf{0.7354} & \multicolumn{1}{c|}{\textbf{-0.0052}} & \multicolumn{1}{c|}{\textbf{0.0209}} & \multicolumn{1}{c|}{\textbf{-0.0135}} & \multicolumn{1}{c|}{\textbf{-0.0576}} & \multicolumn{1}{c|}{\textbf{0.0012}} & \multicolumn{1}{c|}{\textbf{0.03}} & \textbf{-0.0109} & \multicolumn{1}{c|}{\textbf{-0.0215}} & \multicolumn{1}{c|}{\textbf{0.0110}} & \multicolumn{1}{c|}{\textbf{-0.0113}} & \multicolumn{1}{c|}{\textbf{-0.0576}} & \multicolumn{1}{c|}{\textbf{0.0012}} & \multicolumn{1}{c|}{\textbf{0.03}} & \textbf{(0.0157)} \\ \hline
\multicolumn{1}{|c|}{\multirow{2}{*}{\textbf{WMA AccWts}}} & \textbf{Accuracy} & \multicolumn{1}{c|}{\textbf{16.00\%}} & \multicolumn{1}{c|}{\textbf{30.99\%}} & \multicolumn{1}{c|}{\textbf{25.58\%}} & \multicolumn{1}{c|}{\textbf{23.81\%}} & \multicolumn{1}{c|}{\textbf{21.64\%}} & \multicolumn{1}{c|}{\textbf{0.05}} & \textbf{23.60\%} & \multicolumn{1}{c|}{\textbf{24.55\%}} & \multicolumn{1}{c|}{\textbf{25.49\%}} & \multicolumn{1}{c|}{\textbf{31.03\%}} & \multicolumn{1}{c|}{\textbf{25.53\%}} & \multicolumn{1}{c|}{\textbf{22.22\%}} & \multicolumn{1}{c|}{\textbf{0.03}} & \textbf{25.76\%} & \multicolumn{1}{c|}{\textbf{18.35\%}} & \multicolumn{1}{c|}{\textbf{27.78\%}} & \multicolumn{1}{c|}{\textbf{21.43\%}} & \multicolumn{1}{c|}{\textbf{22.62\%}} & \multicolumn{1}{c|}{\textbf{24.03\%}} & \multicolumn{1}{c|}{\textbf{0.03}} & \textbf{22.84\%} \\ \cline{2-23} 
\multicolumn{1}{|c|}{} & \textbf{Utility} & \multicolumn{1}{c|}{\textbf{-0.0700}} & \multicolumn{1}{c|}{\textbf{0.0845}} & \multicolumn{1}{c|}{\textbf{0.0465}} & \multicolumn{1}{c|}{\textbf{0.0119}} & \multicolumn{1}{c|}{\textbf{0.0597}} & \multicolumn{1}{c|}{\textbf{0.06}} & \textbf{0.0265} & \multicolumn{1}{c|}{\textbf{0.0636}} & \multicolumn{1}{c|}{\textbf{0.0588}} & \multicolumn{1}{c|}{\textbf{0.1839}} & \multicolumn{1}{c|}{\textbf{-0.0426}} & \multicolumn{1}{c|}{\textbf{0.0694}} & \multicolumn{1}{c|}{\textbf{0.08}} & \textbf{0.0667} & \multicolumn{1}{c|}{\textbf{-0.0734}} & \multicolumn{1}{c|}{\textbf{0.0000}} & \multicolumn{1}{c|}{\textbf{-0.0595}} & \multicolumn{1}{c|}{\textbf{-0.1190}} & \multicolumn{1}{c|}{\textbf{0.0714}} & \multicolumn{1}{c|}{\textbf{0.07}} & \textbf{-0.0361} \\ \hline
\multicolumn{1}{|c|}{\multirow{2}{*}{\textbf{WMA UtilWts}}} & \textbf{Accuracy} & \multicolumn{1}{c|}{\textbf{20.00\%}} & \multicolumn{1}{c|}{\textbf{38.03\%}} & \multicolumn{1}{c|}{\textbf{28.26\%}} & \multicolumn{1}{c|}{\textbf{27.38\%}} & \multicolumn{1}{c|}{\textbf{25.97\%}} & \multicolumn{1}{c|}{\textbf{0.07}} & \textbf{27.93\%} & \multicolumn{1}{c|}{\textbf{25.47\%}} & \multicolumn{1}{c|}{\textbf{28.17\%}} & \multicolumn{1}{c|}{\textbf{27.17\%}} & \multicolumn{1}{c|}{\textbf{21.28\%}} & \multicolumn{1}{c|}{\textbf{24.43\%}} & \multicolumn{1}{c|}{\textbf{0.03}} & \textbf{25.30\%} & \multicolumn{1}{c|}{\textbf{20.34\%}} & \multicolumn{1}{c|}{\textbf{29.17\%}} & \multicolumn{1}{c|}{\textbf{21.74\%}} & \multicolumn{1}{c|}{\textbf{16.67\%}} & \multicolumn{1}{c|}{\textbf{25.66\%}} & \multicolumn{1}{c|}{\textbf{0.05}} & \textbf{22.71\%} \\ \cline{2-23} 
\multicolumn{1}{|c|}{} & \textbf{Utility} & \multicolumn{1}{c|}{\textbf{0.0000}} & \multicolumn{1}{c|}{\textbf{0.1408}} & \multicolumn{1}{c|}{\textbf{0.1087}} & \multicolumn{1}{c|}{\textbf{0.0238}} & \multicolumn{1}{c|}{\textbf{0.1169}} & \multicolumn{1}{c|}{\textbf{0.06}} & \textbf{0.0780} & \multicolumn{1}{c|}{\textbf{-0.0283}} & \multicolumn{1}{c|}{\textbf{0.0563}} & \multicolumn{1}{c|}{\textbf{0.1196}} & \multicolumn{1}{c|}{\textbf{-0.0745}} & \multicolumn{1}{c|}{\textbf{0.0840}} & \multicolumn{1}{c|}{\textbf{0.08}} & \textbf{0.0314} & \multicolumn{1}{c|}{\textbf{-0.0847}} & \multicolumn{1}{c|}{\textbf{0.0417}} & \multicolumn{1}{c|}{\textbf{-0.0326}} & \multicolumn{1}{c|}{\textbf{-0.1905}} & \multicolumn{1}{c|}{\textbf{0.0461}} & \multicolumn{1}{c|}{\textbf{0.10}} & \textbf{-0.0440} \\ \hline
\end{tabular}%
}
\end{table*}

\section{Related Works}

Recent studies have applied reinforcement learning techniques to trading using price signals alone. Deep Reinforcement Learning in Trading~\cite{zhang2020deep} adopts a formulation similar to Deep Momentum Networks~\cite{lim2019enhancing} and `Momentum Transformer'~\cite{wood2022tradingmomentumtransformerintelligent}, where the trading problem is approached as a task of recommending positions (1 for long, -1 for short, 0 for neutral). However, ~\cite{zhang2020deep} incorporates more advanced reinforcement learning algorithms such as policy-gradient, actor-critic, and deep-Q-learning. The use of deep learning models, particularly CNNs, for predicting financial returns based on image representations of financial data has been explored by ~\cite{vasant_cnn} and \cite{jpm_visionpaper}. While \cite{vasant_cnn} transforms level-2 data (limit-order book) into images, ~\cite{jpm_visionpaper} directly utilizes price series images. Most recently, ~\cite{harini2023neuro} combined meta-reinforcement learning (RL2 algorithm ~\cite{duan2016rl2fastreinforcementlearning}) with automatically learned logical features, including temporal ones, through Inductive Logic Programming (ILP). This novel approach brings together meta-learning and symbolic reasoning to enhance reinforcement learning performance in trading scenarios.

\section{Conclusion and Future Work}
We have presented experiments using weighted-majority ensembles of eight different deep-learning based approaches for predicting ten candle returns
in a realistic setting of intra-day trading equity data. We have used both accuracy as well as the utility metric, which is a proxy for 
potential profitability, to score and re-weight models resulting in two different WMA algorithms. Our results indicate that using WMA results
in better accuracy than the average of all the individual models as well as potential overall profitability even when individual models may not be
profitable consistently. 

We submit that using WMA ensembles in the manner we have described is potentially useful for combining different quantitative trading models
in a dynamic manner as well as for rewarding model builders in a competitive scenario for near real-time intra-day trading, as envisaged in 
our system Numin, inspired by the Numerai competition.

\bibliographystyle{ACM-Reference-Format}
\bibliography{a_my-references}

\end{document}